\input{aipcheck}

\documentclass[
    ,final            
  ]
  {aipproc}

\layoutstyle{6x9}

\begin{document}

\title{Two-Photon Exchange in (Semi-)Inclusive DIS}


\author{\underbar{M. Schlegel}}{
  address={Theory Center, Jefferson Lab, Newport News, VA 23606, USA}
}

\author{A. Metz}{
  address={Department of Physics, Temple University, Philadelphia, PA 19122, USA}
}

\begin{abstract}
In this note\footnote{Authored by Jefferson Science Associates, LLC under U.S. DOE 
Contract No. DE-AC05-06OR23177. The U.S. Government retains a non-exclusive, paid-up, 
irrevocable, world-wide license to publish or reproduce this manuscript for 
U.S. Government purposes.} we consider effects of a Two-Photon Exchange (TPE) in 
inclusive DIS and semi-inclusive DIS (SIDIS). 
In particular, transverse single spin asymmetries are generated in inclusive DIS if 
more than one photon is exchanged between the lepton and the hadron. 
We briefly summarize the TPE in DIS in the parton model and extend our approach to 
SIDIS, where a new leading twist $\sin(2\phi)$ contribution to the longitudinal beam 
spin asymmetry shows up.
Possible TPE effects for the Sivers and the Collins asymmetries in SIDIS are 
power-suppressed.
\end{abstract}

\maketitle


\section{Two-Photon Exchange in Inclusive DIS}
It was already argued in the 60s that, due to time-reversal invariance, single spin 
asymmetries (SSAs) are forbidden in inclusive deep-inelastic lepton scattering off 
nucleons if one considers only a One-Photon Exchange between the lepton and the nucleon 
(Christ-Lee theorem~\cite{Christ:1966zz}). 
Early experiments confirmed this statement and found transverse single target spin 
asymmetries in DIS being consistent with zero at the percent 
level~\cite{Chen:1968mm,Rock:1970sj}. 
Measurements of this observable have been repeated very recently at HERMES with higher 
precision (of the order $10^{-3}$), and the effect is still compatible with 
zero~\cite{DeNardo:2008}. 
An ongoing Hall A experiment at Jefferson Lab even plans to measure the target SSA
in DIS with an accuracy of the order $10^{-4}$~\cite{Jiang:2007}.
 
\begin{figure}[t]
\includegraphics[height=2.5cm,width=12cm]{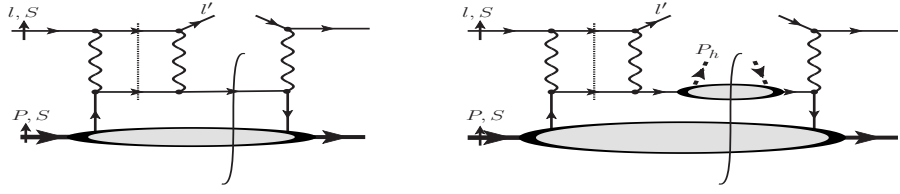}
\caption{Two-Photon Exchange in the parton model.
 Merely the coupling of the two photons to the same quark is considered. 
 Also the complex conjugate diagrams (not shown here) have to be taken into account.
\textbf{Left Panel}: TPE in inclusive DIS.
\textbf{Right Panel}: TPE in semi-inclusive DIS.\label{TPEDIS}}
\end{figure}

If one considers DIS beyond the One-Photon Exchange approximation, transverse SSAs 
may well exist~\cite{Christ:1966zz,Metz:2006pe,Afanasev:2007ii}. 
TPE can generate a nontrivial phase in the amplitude of the process --- a necessary 
condition for a SSA --- and therefore provide a nonzero single spin correlation of 
the form $\epsilon^{\mu \nu \rho \sigma}S_\mu P_\nu l_\rho l^\prime_\sigma$ 
(cf.~Fig.~\ref{TPEDIS}).
The transverse SSAs generated by TPE were calculated in the naive parton model in
Ref.~\cite{Metz:2006pe} by assuming that both photons couple to the same 
quark (Fig.~\ref{TPEDIS}). 
The observables were considered in the Bjorken limit where the momentum transfer 
to the nucleon, $q^2=(l-l^\prime)^2\equiv -Q^2$, is large, and power corrections in 
$1/Q$ were neglected. 
In the following we will briefly discuss this calculation.

Both the transverse target and lepton beam SSA receive their leading contributions in 
$\alpha$ --- the electromagnetic fine structure constant --- from an interference 
of One-Photon Exchange DIS amplitudes and Two-Photon Exchange amplitudes 
(Fig.~\ref{TPEDIS}). 
In the parton model for DIS one considers the hard electromagnetic interaction of 
a quasi-free parton with the lepton and treats the other partons as spectators 
(impulse approximation). 
We assume that, for the particular observables we are interested in here, the 
dominant contribution (in the sense of a twist expansion) of TPE is picked out if 
one works with a single active parton.
This approximation is represented in Fig.~\ref{TPEDIS}. 
The two-photon box diagram is the only QED diagram in the parton model which carries 
an imaginary part. 
Other radiative corrections of the same order in $\alpha$ such as bremsstrahlung 
effects are purely real and hence cannot give rise to transverse SSAs.

The result for the transverse lepton beam SSA, calculated from the left graph in 
Fig.~\ref{TPEDIS} (and its complex conjugate counterpart), reads~\cite{Metz:2006pe}
\begin{equation}
A_{TU}(x_B,y,\phi_s) = \alpha \frac{m_l}{2Q} |\vec{S}_T|\sin(\phi_s) 
\frac{y^2\sqrt{1-y}}{1-y+\frac{1}{2}y^2}
\frac{\sum_q e_q^3 f_1(x_B)}{\sum_q e_q^2 f_1(x_B)}
\label{ATU},
\end{equation}
where $x_B=Q^2/(2P\cdot q)$ and $y=(P\cdot q)/(P\cdot l)$ denote the common DIS 
variables, $e_q$ the quark charge, $|\vec{S}_T|$ the transverse polarization vector 
of the lepton, and $\phi_s$ the angle of this vector with respect to the lepton plane. 
The ordinary unpolarized parton distribution of a quark flavor $q$ is represented by 
$f_1^q$. 
We point out that divergent terms, which appear at intermediate steps of the 
calculation and can be regulated by a photon mass, cancel in the final 
result~(\ref{ATU}).
Since the asymmetry~(\ref{ATU}) is not only proportional to $\alpha\simeq 1/137$ but 
also to the lepton mass $m_l$ one can expect rather small effects. 
In fact, by assuming u-quark dominance in a proton target and an electron beam one 
readily estimates this asymmetry to be of the order $10^{-6}$. 
Asymmetries roughly two-hundred times larger can be generated by a muon beam. 
One might also speculate about enhanced results from effects beyond the naive 
parton model. 
In Refs.~\cite{Afanasev:2004hp,Afanasev:2004pu,Borisyuk:2005rj,Gorchtein:2005yz} 
(double) logarithms of the type $\log(Q^2/m_l^2)$ were advocated in connection with 
the transverse beam SSA in elastic lepton-nucleon scattering.
Such logarithmic terms might also increase the beam SSA in DIS considered here.
However, further work is required in order to decide whether and how precisely such 
effects show up for the DIS case. 

We now turn to the parton model result for the transverse target SSA in DIS.
The contribution of the left diagram in Fig.~\ref{TPEDIS} (and its complex 
conjugate counterpart) reads~\cite{Metz:2006pe}
\begin{equation}
A_{UT}(x_B,y,\phi_s) = \alpha \frac{x_B M}{2Q} 
\frac{y(1-y)\sqrt{1-y}}{1-y+\frac{1}{2}y^2} |\vec{S}_T| \sin(\phi_s)
\Big( \ln\frac{Q^2}{\lambda^2} + \textrm{finite} \Big) 
\frac{\sum_q e_q^3 g_T^q(x_B)}{\sum_q e_q^2 f_1^q(x_B)},
\label{TSSA}
\end{equation}
where $M$ denotes the nucleon mass, and $g_T$ a higher twist parton distribution. 
Note that a small photon mass $\lambda$ has been introduced in order to get a 
finite result.
The divergence for $\lambda \to 0$ signals a violation of electromagnetic gauge 
invariance in the parton model (for the particular observable $A_{UT}$).
It was suggested that gauge invariance might be restored by considering 
contributions beyond the naive parton model such as multiparton 
correlations~\cite{Metz:2006pe}. 
Since the SSA~(\ref{TSSA}) is a twist-3 observable one indeed expects contributions 
from quark-gluon correlations. 
Whether the inclusion of such terms provides a finite result remains to be seen.
Another candidate for the restoration of electromagnetic gauge invariance are 
multiquark correlations, i.e. configurations where the two photons couple to two 
different quarks. 

Focusing on the finite contributions in~(\ref{TSSA}) one expects the asymmetry 
to be at most of the order $10^{-3}$ (for $x \simeq 0.1$).
(A very rough estimate of $A_{UT}$, where one just considers the suppression due 
to $\alpha$, gives an effect of the order $10^{-2}$.)
However, because of the divergence, a reliable numerical result can hardly be 
extracted from Eq.~(\ref{TSSA}).
Additional contributions beyond the naive parton model not only are supposed to 
cancel the divergence but also to yield finite terms which have to be taken into 
account as well.
It is also worthwhile to mention that a quark-mass effect to $A_{UT}$ --- 
a particular contribution beyond the parton model --- was studied in detail in 
Ref.~\cite{Afanasev:2007ii}.
Such an effect is proportional to the quark transversity 
distribution and leads, according to~\cite{Afanasev:2007ii}, to an asymmetry of 
the order $10^{-4}$.
In addition, we note that a diquark spectator model calculation for the neutron 
also predicts $A_{UT} \simeq 10^{-4}$~\cite{Afanasev:2009}.

\section{Two-Photon Exchange in Semi-inclusive DIS}
It is straightforward to extend the naive parton model calculation of 
Ref.~\cite{Metz:2006pe} to SIDIS in the kinematical regime where the transverse 
momentum $P_{h\perp}$ of the produced hadron is much smaller than the momentum 
transferred to the nucleon, $P_{h\perp}\ll Q$. 
For this kinematics, observables can be described in the parton model in terms 
of transverse momentum dependent (TMD) parton distributions and fragmentation 
functions. 
Like in inclusive DIS, TPE in general also generates SSAs in SIDIS. 
In the following we discuss results for the SSAs calculated in the naive parton 
model. 
The contribution of the right graph in Fig.~\ref{TPEDIS} reads for a longitudinally 
polarized lepton beam and a produced pion,
\begin{equation}
 A_{LU}^{\sin(2\phi)} = 
\alpha \frac{y\Big(1+\frac{2-y}{1-y}\ln y\Big)}{1-y+\frac{1}{2}y^2} 
\sin(2\phi) \frac{\sum_q e_q^3 \mathcal{C} 
\Big[\frac{2(\vec{h}\cdot \vec{k}_T)(\vec{h}\cdot \vec{p}_T)-\vec{k}_T\cdot \vec{p}_T}
{2Mm_\pi}h_1^{\perp q}H_1^{\perp q}\Big]}
{\sum_q e_q^2 \mathcal{C}\Big[f_1^q D_1^q \Big]}.
\label{ALU}
\end{equation}
The angle $\phi$ represents the angle between the lepton and the production plane 
(in the conventions of Ref.~\cite{Bacchetta:2006tn}), while $D_1$ denotes the 
unpolarized TMD fragmentation function, $h_1^\perp$ the Boer-Mulders 
function~\cite{Boer:1997nt}, and $H_1^\perp$ the Collins function~\cite{Collins:1992kk}. 
We also use the unit-vector $\vec{h} \equiv - \vec{P}_{h\perp} / |\vec{P}_{h\perp}|$.
The symbol $\mathcal{C}[\dots]$ denotes a convolution in the transverse momentum space 
($w$ being some kinematical prefactor),
\begin{equation}
\mathcal{C}[w\,f\,D] = \int d^2k_T d^2p_T \, 
\delta^{(2)}(\vec{k}_T-\vec{p}_T-\vec{P}_{h\perp}/z_h) \,
w(\vec{k}_T,\vec{p}_T)f^q(x_B,\vec{k}_T^2)D^q(z_h,\vec{p}_T^2).
\end{equation}
Both TMD functions in Eq.~(\ref{ALU}) are so-called (naive) time-reversal odd 
entities for which initial/final state interactions play an important role. 
We assume that the interplay between initial/final state interactions and TPE 
does not spoil the QCD-factorization in terms of TMD correlation functions.

Like Eq.~(\ref{ATU}), the SSA in~(\ref{ALU}) is free of divergences.
Furthermore, it is a leading twist $\sin(2\phi)$ modulation of the beam spin 
asymmetry that is absent in a model-independent decomposition of the SIDIS 
cross section into structure functions assuming  
One-Photon Exchange~\cite{Kotzinian:1994dv,Diehl:2005pc,Bacchetta:2006tn}. 
This means that the asymmetry (\ref{ALU}) is generated by TPE only, and 
possible $\sin(2\phi)$ modulations from QCD radiative corrections in the 
One-Photon Exchange approximation are ruled out. 
A similar effect has also been discussed in~\cite{Pascalutsa:2005es} for  
TPE in electro-excitation of a $\Delta$ resonance. 
Since Eq.~(\ref{ALU}) is proportional to $\alpha$ and since one expects the 
Boer-Mulders effect in unpolarized SIDIS from a $\cos(2\phi)$ modulation to be 
of the order of a few percent (see, e.g., the model calculation~\cite{Gamberg:2007wm})
one can roughly estimate the asymmetry in~(\ref{ALU}) to be of the order 
$10^{-4} - 10^{-3}$. 
At this point one might again speculate about possible logarithmic enhancements 
as discussed above for the transverse beam SSA in inclusive DIS. 

We also calculated the target asymmetries $A_{UL}$ and $A_{UT}$, where we again
encounter divergences like in the case of the transverse target SSA in DIS.
The comments on this feature which we made above apply here, too.
Interestingly, in the case of the target SSAs, TPE provides only twist-3 
contributions to azimuthal modulations like $\sin \phi$ for $A_{UL}$, as well 
as $\sin \phi_s$ and $\sin (2\phi-\phi_s)$ for $A_{UT}$. 
Such twist-3 modulations are already present in the One-Photon Exchange 
approximation~\cite{Bacchetta:2006tn}. 
In particular, we find that for $A_{UT}$ TPE corrections to the modulations
$\sin (\phi-\phi_s)$ and $\sin (\phi+\phi_s)$ which are typically analyzed in
terms of the (leading twist) Sivers and Collins effect, respectively, appear at 
most at the level of twist-4.



\bibliographystyle{aipproc}   

\bibliography{Referenzen}

\end{document}